\documentclass[twocolumn]{article}

\begin{document}
\title{Information extraction and quantum state distortions
in continuous variable quantum teleportation}

\author{Holger F. Hofmann$^{a}$, Toshiki Ide$^{b}$, 
Takayoshi Kobayashi$^{b}$,
and Akira Furusawa$^{c}$\\
$^{a}$CREST, Japan Science and Technology Corporation (JST),\\
Research Institute for Electronic Science,
Hokkaido University, Sapporo 060-0812, Japan\\
$^{b}$
Department of Physics, Faculty of Science, University of Tokyo,\\
7-3-1 Hongo, Bunkyo-ku, Tokyo 113-0033, Japan\\
$^{c}$
Department of Applied Physics, Faculty of Engineering, 
University of Tokyo,\\ 7-3-1 Hongo, Bunkyo-ku, Tokyo 113-8656, 
Japan}   

\date{}

\maketitle

\abstract{We analyze the loss of fidelity in continuous variable
teleportation due to non-maximal entanglement. It is shown that
the quantum state distortions correspond to the measurement 
back-action of a field amplitude measurement}

\section{Introduction}

Quantum teleportation transfers a quantum state to a remote location
using shared entanglement and classical communication \cite{Ben93}. 
Ideally, this
procedure does not change the transmitted state at all, even though 
classical information is obtained in an irreversible measurement.
This is only possible if the classical information is completely
independent of the teleported state. In the case of continuous 
variable quantum teleportation \cite{Ctele}, only non-maximal 
entanglement is available. As a result, the classical information 
obtained in the measurement does depend on the input
state, and a corresponding measurement back-action is observed in the
output. In the following, this limitation of fidelity in continuous 
variable teleportation is discussed using the recently introduced
transfer operator formalism \cite{Hof00}. 
\vspace{\fill}

\section{Continuous variable\\ teleportation}
In continuous variable teleportation, an unknown input state 
$\mid \psi_{\mbox{in}}\rangle$ of the input field $\hat{a}$ 
is transfered by a precise measurement of the field difference 
$\hat{a}-\hat{r}^\dagger=\beta$ between $\hat{a}$ and a reference field
$\hat{r}$. 
The reference field $\hat{r}$ is entangled with the remote 
field $\hat{b}$. This entanglement is obtained by squeezing the vacuum
to suppress the fluctuations of $\hat{b}-\hat{r}^\dagger\approx 0$ below 
the standard quantum limit.
Therefore, the measurement result $\beta$ is approximately equal to the 
field difference between the unknown input field $\hat{a}$ and the output 
field $\hat{b}$. The original quantum state of the input field can then be 
restored by $\hat{b} + \beta \approx \hat{a}$.

The most serious technical limitation of this teleportation scheme
is the amount of squeezed state entanglement available. At present,
it seems unrealistic to assume a noise suppression of more than 10 dB.
While it may be possible to raise this limit in the future, maximal
entanglement would require the unrealistic limit of infinite squeezing.
Non-maximal entanglement is therefore a fundamental feature of continuous
variable teleportation. The quantum state of non-maximal squeezed state
entanglement can be formulated in the photon number basis as
\\
\begin{equation}
\mid EPR (q) \rangle = 
\sqrt{1-q^2}\sum_n q^n \mid n \rangle \otimes \mid n \rangle.
\end{equation}
The entanglement parameter $q$ is related to the squeezing factor $s$
by $q=(1-s)/(1+s)$. For example, a noise suppression by one half (3 dB)
would correspond to $q=1/3$. 

With this definition of the initial state, it is possible to describe 
how the measurement of $\hat{a}-\hat{r}^\dagger=\beta$ conditions the
output state of $\hat{b}$. Using the properly normalized eigenstates
of $\hat{a}-\hat{r}^\dagger$ \cite{norm}, the projection reads
\\
\begin{equation}
\begin{array}{cccccc}
\multicolumn{6}{l}{\mbox{Initial state}}\\[0.2cm]
\sqrt{1-q^2}\sum_n q^n & \mid \psi_{\mbox{in}} \rangle & \otimes 
& \mid n \rangle & \otimes & \mid n \rangle \\[0.3cm]
\multicolumn{6}{l}{\mbox{Measurement projection}}\\[0.2cm]
\frac{1}{\sqrt{\pi}} \sum_n & \langle n \mid \hat{D}(-\beta) 
& \otimes & \langle n \mid& & \\[0.3cm]
\multicolumn{6}{l}{\mbox{Conditional output state}}\\[0.2cm]
\sqrt{\frac{1-q^2}{\pi}} \sum_n q^n & & 
\multicolumn{3}{c}{\hspace{-2cm}\langle n \mid 
\hat{D}(-\beta)\mid \psi_{\mbox{in}}\rangle}
& \mid n \rangle.\\[0.5cm]
\end{array}
\end{equation}
\\[0.2cm]
For $q \to 1$, the conditional output state is equal to the
displaced input state $\hat{D}(-\beta)\mid~\psi_{\mbox{in}}~\rangle$.
The effect of $q<1$ reduces the contributions of states with
high photon numbers.

In the final step of quantum teleportation, the displacement is 
reversed by modulating the output field $\hat{b}$. This 
modulation is proportial to $\beta$ and can be described by a
displacement operator $\hat{D}(g\beta)$, where the gain factor $g$
permits an amplification or attenuation of the output amplitude.
The process of continuous variable quantum teleportation can then
be described by a transfer operator, such that both the probability
distribution $P(\beta)$ of the measurement results and the normalized
conditional output states $\mid \psi_{\mbox{out}}(\beta)\rangle$ 
are described by
\\
\[
\sqrt{P(\beta)} \mid \psi_{\mbox{out}}(\beta)\rangle = 
\hat{T}(\beta) \mid \psi_{\mbox{in}}\rangle 
\]
with
\begin{eqnarray}
\lefteqn{\hat{T}(\beta) =}\nonumber \\[0.3cm]
&&\sqrt{\frac{1-q^2}{\pi}} \sum_{n=0}^\infty\; q^n\;
\hat{D}(g \beta)\mid n \rangle \langle n \mid \hat{D}(-\beta).
\nonumber \\
\end{eqnarray}
The transfer operator $\hat{T}(\beta)$ establishes the general
relationship between the measurement information $\beta$ obtained
in the teleportation and the conditional quantum state distortions
caused by this measurement for arbitrary input states. In particular, 
the information obtained about the unknown input state 
$\mid \psi_{\mbox{in}}\rangle$ is characterized by a positive operator
valued measure given by
\\
\[
P(\beta) = \langle \psi_{\mbox{in}} \mid \hat{T}^\dagger(\beta)
\hat{T}(\beta)\mid \psi_{\mbox{in}}\rangle 
\]
with
\begin{eqnarray}
\label{eq:povm}
\lefteqn{\hat{T}^\dagger(\beta)\hat{T}(\beta) =} \nonumber \\[0.3cm] &&
\frac{1-q^2}{\pi} \sum_{n=0}^\infty\; q^{2n}\;
\hat{D}(\beta) \mid n \rangle \langle n \mid \hat{D}(-\beta).
\nonumber \\
\end{eqnarray}
The eigenvalues of this positive operator valued measure are the 
displaced photon number states $\hat{D}(\beta) \mid~n~\rangle$.
In phase space, these displaced photon number states can be associated
with concentric circles of radius $\sqrt{n+1/2}$ around $\beta$.
The higher $n$, the greater the difference between the actual field 
value and $\beta$. $\hat{T}(\beta)$ therefore describes a finite 
resolution measurement of the complex field amplitude $\hat{a}$. 

\section{Coherent state teleportation}
The properties of the transfer operator are best illustrated by
applying it to typical input states. For a coherent
state input $\mid \alpha \rangle$, the quantum teleportation 
process is characterized by
\\
\begin{eqnarray}
\label{eq:cs}
\lefteqn{\hat{T}(\beta) \mid \alpha \rangle =}
\nonumber \\[0.3cm] &&
\sqrt{\frac{1-q^2}{\pi}} 
\exp\left(-(1-q^2)\frac{|\alpha-\beta|^2}{2}\right) \nonumber \\ 
&\times&
\exp\left((1-g q)\frac{\alpha\beta^*-\beta\alpha^*}{2}\right) 
\nonumber \\ &&
\times
\mid q \alpha +(g-q) \beta \rangle. \nonumber \\
\end{eqnarray}
This result consists of a probability factor $\sqrt{P(\beta)}$
describing a Gaussian probability distribution centered around
$\beta=\alpha$, a phase factor important only if a superposition
of coherent states is considered (e.g. cat states or squeezed 
states), and finally the modified coherent state, with an 
attenuated original amplitude of $q \alpha$ and a measurement 
dependent displacement of $(g-q)\beta$. 

Note that equation (\ref{eq:cs}) may be applied to any input state
if that state is written as a superposition of coherent states. 
The phase factor is then crucial in determining the coherence of
the output. For a simple coherent state, however, the distortions 
of the output state are best characterized by defining the measurement 
fluctuation $\phi=\beta-\alpha$. The probability distribution over
$\phi$ is then given by a Gaussian centered around $\phi=0$, and
the output statistics are described by
\\
\begin{eqnarray}
P(\phi) \hspace{0.4cm} &=& 
\frac{1-q^2}{\pi}\exp\left(-(1-q^2)|\phi|^2\right)
\nonumber \\
\mid \psi_{\mbox{out}}(\phi)\rangle &=& \hspace{0.6cm}
\mid g \alpha + (g-q) \phi \rangle. \nonumber \\
\end{eqnarray}
The correlation between the measurement fluctuation $\phi$ and the
output amplitude $g\alpha + (g-q)\phi$ is given by the gain dependent
factor $g-q$. In particular, $g>q$ indicates a positive correlation
between the measurement fluctuation and the output amplitude, while
$g<q$ indicates a negative correlation. In the special case of
$g=q$, the output amplitude does not depend on the measurement result.
At this gain condition, continuous variable quantum teleportation
simply attenuates the coherent state to an amplitude of $q\alpha$.
As pointed out by Polkinghorne and Ralph \cite{Pol99}, this 
situation corresponds to the attenuation of the signal at a beam 
splitter. Our formalism allows a generalization of this analogy to
back action evasion measurements using feedback compensated beam 
splitters \cite{Hof01}.

\section{Photon number state\\ teleportation}
It is possible to identify the beam splitter analogy more directly
by examining the effects of the Transfer operator on the creation operator
$\hat{a}^\dagger$. For $g=q$,
\\
\begin{eqnarray}
\hat{T}_{g=q}(\beta) \hat{a}^\dagger &=& \left( (1-q^2)\beta^*
+ q \hat{a}^\dagger\right) \hat{T}_{g=q}(\beta).\nonumber \\
\end{eqnarray}
Effectively, $\hat{T}_{g=q}(\beta)$ attenuates $\hat{a}^\dagger$
by a factor of $q$ and replaces the loss with a complex amplitude
of $\sqrt{1-q^2} \beta^*$. The attenuated amplitude can be identified
with the component transmitted by a beam splitter of reflectivity 
$R=1-q^2$ and the $\beta^*$ dependent addition can be interpreted
as the measurement back-action from the reflected parts of the input field.
In general, the teleportation of a photon number state can then be 
described by
\\
\begin{eqnarray}
\lefteqn{\hat{T}(\beta) \frac{1}{\sqrt{n!}} \left(\hat{a}^\dagger\right)^n 
\mid 0 \rangle =} \nonumber \\
&& \hspace{1.2cm} \hat{D}((g-q)\beta) \nonumber \\[0.2cm] &\times&
\sqrt{\frac{1-q^2}{\pi n!}}
\exp\left(-(1-q^2)\frac{|\beta|^2}{2}\right) \nonumber \\ && \times
\left((1-q^2)\beta^*+q \hat{a}^\dagger\right)^n \mid 0 \rangle.
\nonumber \\
\end{eqnarray}
The measurement back-action causes photon losses and introduces coherence
by replacing a component of the creation operators with a complex
amplitude. This corresponds to the loss of photons at a beam splitter
and the measurement back-action of a projection on coherent states, e.g. 
by eight port homodyne detection.

\section{Conclusions}
The transfer operator $\hat{T}(\beta)$ provides a complete description 
of the measurement information extracted and the quantum state distortions
in continuous variable teleportation. The correlations between the errors 
caused and the information obtained correspond to the back-action of
a non-destructive quantum measurement of the coherent field amplitude. 
In particular, the distortions correspond to the attenuation of the
original signal amplitude and a measurement back-action conditioned by
the field information obtained from a measurement of these losses.


\begin{thebibliography}{Xyz00}
\bibitem{norm} Note that the proper normalization for the continuous
variable $\beta$ is the delta function 
$\langle \beta \mid \beta^\prime\rangle = \delta (\beta-\beta^\prime)$.
\end{thebibliography}
\end{document}